\documentstyle[12pt]{article}

\def\GeV{\,{\rm GeV}}
\def\TeV{\,{\rm TeV}}

\def\kpc{\,{\rm kpc}}
\def\Mpc{\,{\rm Mpc}}

\def\eV{{\,\rm eV}}

\def\cmm2{{\,\rm cm^{-2}}}
\def\cm2{{\,{\rm cm}^2}}
\def\cmm3{{\,{\rm cm}^{-3}}}
\def\gcmm3{{\,{\rm g\,cm^{-3}}}}
\def\kms{\,{\rm km\,s^{-1}}}

\def\ga{\mathrel{\mathpalette\fun >}}
\def\fun#1#2{\lower3.6pt\vbox{\baselineskip0pt\lineskip.9pt
  \ialign{$\mathsurround=0pt#1\hfil##\hfil$\crcr#2\crcr\sim\crcr}}}

\begin{document}
\pagestyle{empty}
\begin{center}

\rightline{FERMILAB--Pub--93/357-A}
\rightline{astro-ph/yymmdd}
\rightline{Submitted to {\it Physical Review Letters}}

\vspace{.4in}
{\Large \bf HALO COLD DARK MATTER}\\
\medskip
{\Large \bf AND MICROLENSING}

\vspace{.3in}
Evalyn Gates$^{a,b}$ and Michael S. Turner$^{a,b,c}$\\

\vspace{0.2in}

{\it $^a$Department of Astronomy \& Astrophysics\\
Enrico Fermi Institute, The University of Chicago, Chicago, IL~~60637-1433}\\

\vspace{0.1in}

{\it $^b$NASA/Fermilab Astrophysics Center\\
Fermi National Accelerator Laboratory, Batavia, IL~~60510-0500}\\

\vspace{0.1in}

{\it $^c$Department of Physics\\
Enrico Fermi Institute, The University of Chicago, Chicago, IL~~60637-1433}\\

\end{center}

\vspace{.3in}

\centerline{\bf ABSTRACT}

\medskip

There is good evidence that most of the baryons in the
Universe are dark and some evidence that most of the matter in the Universe is
nonbaryonic with cold dark matter (cdm) being a promising possibility.
We discuss expectations for the abundance of
baryons and cdm in the halo of our galaxy and locally.  We show that
in plausible cdm models the local density of cdm is at least
$10^{-25}\gcmm3$.   We also discuss what one can learn about the
the local cdm density from microlensing of stars in
the LMC by dark stars in the halo and, based upon a suite
of reasonable two-component halo models, conclude that
microlensing is not a sensitive probe of the local cdm density.

\pagestyle{plain}
\setcounter{page}{1}
\newpage

{\it Introduction.}
While the quantity and composition of matter in the Universe is
still not known with certainty,
it is known that:  (i) luminous matter (stars, etc.)
contributes much less than 1\% of critical density;
(ii) based upon primordial nucleosynthesis
baryons contribute between about 1\% and 10\% of critical density;
and (iii) based upon numerous
dynamical measurements the total mass density is {\it at least} 10\% of
critical density \cite{dark}, with several determinations indicating that
it is close to the critical density \cite{omega}.  Thus, there is
overwhelming evidence that most of the matter in the
Universe is ``dark,'' compelling evidence that most of the
baryons are dark, and mounting evidence that most of the matter
is nonbaryonic \cite{mst1}.  If the mean mass density is greater
than about 10\% of critical there are two dark matter problems, the nature
of the baryonic and nonbaryonic dark matter.

The case for a critical Universe with nonbaryonic dark
matter receives further support
from studies of structure formation:  The most successful models
rely upon nonbaryonic dark matter, and the cold-dark matter
models of structure formation (inflation-produced
density perturbations and nonbaryonic dark matter with
negligible velocity dispersion) are
very attractive \cite{structure}.  The best motivated cold
dark matter candidates are an axion of
mass around $10^{-5}\eV$ and a neutralino of mass between $10\GeV$
and $1\TeV$ \cite{cold}, and large-scale experiments are
underway to directly detect the axions or neutralinos
in the halo \cite{expt}.
Needless to say, theoretical expectations for, and observational
information about, the local mass density of
cold dark matter are of great importance.

The flat rotation curves of spiral galaxies
indicate that the luminous, disk shaped portion of a typical spiral
sits in a dark halo that is roughly spherical with
density that decreases as $1/(r^2+a^2)$ and extent that
is undetermined ($r$ is distance from the center of the galaxy
and $a$ is the core radius).
The fact that galactic halos are more spherical and
extended than the luminous parts of spirals
strongly suggests that the dark halo material
has probably not undergone significant dissipation.
For the Milky Way, galactic modeling indicates that the core
radius is between $2\kpc$ and $8\kpc$ and that the halo
density nearby ($r\equiv r_0\simeq 8.5\kpc$) is about
$5\times 10^{-25}\gcmm3$ (to within a factor of two) \cite{halomodel}.
At our position, the halo material supports around $130\kms$ of
the $220\kms$ circular rotation velocity (the various contributions---halo,
disk, etc.---to the rotation velocity add in quadrature.)

In a cold dark matter Universe
there are at least two forms of dark matter, baryons and
cdm particles, and both are expected to contribute to the halo mass density.
The concerns of this {\it Letter}
are twofold: first, the theoretical expectations for the
local cdm density, and second, what one can learn
about the local cdm density from microlensing experiments which
can probe the baryonic component that exists in dark stars
of mass $10^{-6}M_\odot$ to $100M_\odot$ \cite{pac}.  Since the EROS and
MACHO microlensing searches now have candidate
microlensing events \cite{macho}, this is a very timely issue.

\medskip
{\it Expectations.}  Consider a cold dark
matter universe with $\Omega_{\rm cdm} + \Omega_B =1$.  Based upon
primordial nucleosynthesis $0.01h^{-2} \le\Omega_B
\le 0.02h^{-2}$; further, in a critical, matter-dominated Universe
the Hubble constant must be near its lower extreme, $h\sim 0.5$,
in order to accommodate a sufficiently aged Universe
($H_0=100h\kms\Mpc^{-1}$ is the present Hubble constant and
$\Omega_i$ is the fraction of critical density contributed
by species $i$).  This means that
the universal baryon fraction of the matter density
is $f_B = \Omega_B = 0.04-0.1$ and $f_{\rm cdm} = 1 -f_B\sim 0.9 - 0.96$.

(In the two popular variants of cold dark matter, hot + cold dark
matter, where $\Omega_\nu \sim 0.3$, and cold dark matter
+ cosmological constant, where $\Omega_\Lambda \sim 0.8$,
$\Omega_{\rm cdm} \sim 0.2$ and $h\sim 0.7$, the situation
is only quantitatively different.  In the former case,
phase-space considerations
limit the neutrino fraction in galaxies like our own
to be less than about 5\% \cite{nufrac}, and so
the baryonic fraction of matter that can clump
in galaxies, $f_B\simeq \Omega_B/(1-\Omega_\nu)\simeq 0.06-0.14$,
is slightly higher.  In the latter case, the baryonic
fraction of matter, $f_B = \Omega_B/(1-\Omega_\Lambda )
\simeq 0.1-0.2$, is even larger.  All that follows
is applicable to these models by taking account
of the larger value of $f_B$.)

So long as gravity alone shapes the evolution of the Universe
the baryonic fraction of matter remains at
its universal value.  Once dissipative forces (e.g.,
electromagnetic interactions) become
important, baryons can lose energy and become more condensed,
increasing the local baryonic fraction.
The formation of the disks of spiral galaxies provides a good
example:  Through collisional processes
baryons lose their energy,
but not angular momentum, ultimately forming a thin, rotationally
supported disk.  The local mass density of the disk is about
a factor of 20 greater than that of the halo.

In turn we now consider three models
for the formation of the halo of our galaxy, from a very simple
scenario where only gravity is involved to an extreme
scenario where hydrodynamical forces form the galaxy and the cold dark
matter particles are captured subsequently by accretion.

(1) The simplest and most plausible scenario
is one where the halo of our galaxy formed
through the action of gravity alone, after which a small fraction
of the baryons dissipated energy forming the disk.
One then expects an isothermal halo, with density
decreasing as $r^{-2}$, whose extent is dictated by the ``sphere
of influence'' of our galaxy, at most about half the distance to
M31 ($\approx 400\kpc$) \cite{numerical}.  The baryonic fraction
of the halo mass density should be about equal to the universal
value, $f_B \simeq 0.04-0.1$, or smaller, if a larger fraction
of the baryons dissipated their energy and reside in the disk.

(2) Suppose further that a substantial fraction of halo baryons
undergo moderate dissipation (though not enough to
collapse to form a disk), so that the baryonic halo remains
roughly spherical but shrinks in size.  Specifically, assume
that both cdm and baryons exist in isothermal
halos truncated at different radii, $R_B$ and $R_{\rm cdm}=\beta R_B$.
If most of the baryons are in the halo, the
ratio of the halo mass in baryons to that in cdm is $f_B/(1-f_B)
\simeq f_B$, and interior to $R_B$ the ratio of baryons
to cdm is $\beta f_B/(1-f_B) \simeq \beta f_B$.
That is, the local baryon to cdm
ratio is increased by the ratio of their truncation radii.  To be concrete,
if $R_{\rm cdm} \sim 200\kpc$ and $R_B\sim 30\kpc$,
then interior to $30\kpc$ the ratio of baryons to cdm
is about $7f_B \sim 0.3-0.7$.

(3)  Consider a very radical scenario, one where our
galaxy formed by nongravitational forces (e.g., hydrodynamical
shock waves), so that the baryonic part of the galaxy was
assembled first and the cdm halo accreted subsequently.
By using the spherical accretion model \cite{sam} one can estimate the
mass of cdm halo that is added.   Suppose at
time $t_i$ a point mass $M_0$ is placed at the
origin of an otherwise smooth critical Universe comprised of
cold dark matter.  Thereafter, all cdm particles in the Universe
are gravitationally bound to $M_0$ and ultimately cease moving
away (clearly the model is only applicable within the sphere
of influence of $M_0$, say half way to M31.)

According to this model, cdm is accreted in a self-similar way,
with a density profile $\rho \propto r^{-9/4}$
for $r\ll r_*$; at time $t$ the ``turn-around''
radius $r_*(t) = 0.771(GM_0)^{1/3}t^{8/9}/t_i^{2/9}$.
The mass accreted by time $t$ and interior to radius
$r\ll r_*$ is:  $M(r,t) = 1.39(M_0 r)^{3/4}/
G^{1/4}t_i^{1/2}$.  Taking $M_0\sim
10^{11}M_\odot \sim $ (baryonic mass of our galaxy), the turn-around
radius today is around $1\Mpc$, and the mass accreted
within $100\kpc$ is about $7\times 10^{10}M_\odot (1+z_i)^{3/4}h^{1/2}
\sim 10^{11}M_\odot$  for $z_i\sim 1-2$ and $h\sim 0.5$
($z_i$ is the red shift corresponding to time $t_i$).  The density
of cdm particles $8.5\kpc$ from the galactic
center is $0.7\times 10^{-25}\gcmm3 (1+z_i)^{3/4} h^{1/2}
\sim 10^{-25}\gcmm3$ for $z_i\sim 1-2$ and $h\sim 0.5$.
Even in this radical model, the amount of cold dark matter
accreted within $100\kpc$ is about equal to the baryonic mass,
and the local density of cold dark matter is only slightly lower
than the estimates in scenarios (1) and (2).

(One can consider a more extreme version of this
scenario:  suppose that the Milky Way resides in an underdense
region of the Universe, which, in the absence of the point mass $M_0$, behaves
like a small portion of an $\Omega_0 < 1$ Universe.  In this
case, the total cdm mass accreted is $(1+z_i)\Omega_0 M_0/(1-\Omega_0)$,
again comparable to $M_0$, and so the previous comments apply.)

\medskip
{\it Microlensing.}  Now we turn to what
one can hope to learn experimentally from the study of
microlensing of stars in the LMC by dark stars in the halo
of our galaxy.  Such experiments can only ``detect'' halo baryons
if they exist in the form of $10^{-6}M_\odot$ to $100M_\odot$ dark stars.
Little is known about the kind of objects halo baryons would form
(in part, because of our poor understanding of star formation
in general).  The strongest statements that can be made concern in what
form halo baryons cannot exist and lead to
the suggestion that halo baryons are likely to
be stars of mass $10^{-3}M_\odot$
to $0.1M_\odot$ \cite{halobaryons}, a range that can
be probed by microlensing.

In modeling a two-component halo we assume that:  (i) baryons and cdm exist in
separate, spherically-symmetric isothermal halos,
with core radii $a_B$ and $a_{\rm cdm}$ and density profiles,
\begin{equation}
\rho_i(r) = \rho_{{\rm local,}i}\,\left( {a_i^2+r_0^2\over a_i^2 +r^2}\right) ,
\end{equation}
where $\rho_{{\rm local},i}$ is the local density;
(ii) the core radii are between $2\kpc$ and $8\kpc$;  (iii) halo baryons
are dark stars of mass $M$.  Both the baryonic
and cdm halos play a role in determining the galactic rotation
curve, but of course only the baryonic halo determines the microlensing rate.

Next we compute the
microlensing rate for stars in the Large Magellanic Cloud (LMC)
as a function of the {\it assumed} local density of cdm
for a suite of reasonable halo models (see below).  That rate depends upon
the distribution of baryonic matter in the halo and is given by
\begin{equation} \label{eq:micro}
\Gamma (\rho_{\rm local,cdm},\, a_i) = \omega_0 u_T
 \int_0^{x_{\rm max}} {dx \sqrt{x(1-x)}\over A^\prime
+ B x + x^2},
\end{equation}
where $\omega_0 = \sqrt{8\pi GL^3 \over 3Mc^2}v_H A^\prime\rho_{\rm local, B}$,
$A^\prime = (a_B^2+r_0^2)/L^2$, and
$B=-2(r_0/L)\cos b \cos l$, $v_H$
is the halo-velocity dispersion, and $L\simeq 50\kpc$,
$b=-33^\circ$, and $l=281^\circ$ are respectively the
distance, and galactic latitude
and longitude of the LMC.  The quantity $u_T$, the
threshold impact parameter in units of the Einstein radius,
is set by the minimum amplification that can be detected;
e.g., $u_T=1$ corresponds to
a amplification threshold of 1.34 which is typical of current
searches \cite{macho}.  The quantity $x_{\rm max}$
is the lesser of 1 and the distance to the edge of the
baryonic halo along the line of sight to the LMC in units of $L$.

We compare all microlensing rates to a fiducial model,
a baryons-only, ``best fit'' halo model
with $a_B=5\kpc$, $v_H=270\kms$, normalized
to have rotation velocity of $220\kms$ at
our position.  The microlensing rate for this model is $\Gamma_0 = 1.66\times
10^{-6}u_T/\sqrt{M/M_\odot}$ events yr$^{-1}$.  For further
discussion of microlensing we refer the reader to Ref.~\cite{kg}.

Our suite of models was constructed as follows:
For each assumed value of the local cdm density, we allow
the core radii to vary separately between $2\kpc$ and $8\kpc$; the value of
$\rho_{\rm local, B}$ is determined by constraining the
rotation velocity at our position to be $v_c (r_0)
= 220\pm 10 \kms$,
\begin{eqnarray*}
\rho_{\rm local,B} & = & {(a_B^2+ r_0^2)^{-1}
\over 1-(a_B/r_0) \tan^{-1}(r_0/a_B)} \nonumber \\
& \times & \left[ {v^2\over 4\pi G} -
\rho_{\rm local,cdm}(a_{\rm cdm}^2+r_0^2)
[1-(a_{\rm cdm}/r_0)\tan^{-1} (r_0/a_{\rm cdm})]\right]
\end{eqnarray*}
where $v(r_0) \simeq 130 \pm 17\kms$ is the portion of the local
rotational velocity that is supported by the halo.
To ensure that a given model is ``reasonable'' we construct the
rotation curve; in so doing we also take into
account the contributions of the disk, bulge, and central
components of the galaxy by using results from Ref.~\cite{halomodel}.
For our ``criterion of reasonableness'' we follow
Ref.~\cite{gould}:  the relative difference of the maximum
and minimum rotation velocities over the interval
$3 \kpc$ to $18 \kpc$ must be less than 14\%
(most of the models pass this test; see Fig.~2).
In the limit of a single halo component, the local halo density
in our models is $3-7\times 10^{-25}\gcmm3$, consistent
with previous estimates \cite{halomodel}.

The range of microlensing rates for our suite of reasonable
models is shown in Fig.~1 as a function of the local cdm
density.   The microlensing
rate is relatively insensitive to the local cdm density---the
range of cdm densities consistent with a given $\Gamma$ spans
about a factor of 3---and is relatively sensitive
to the halo model parameters---for fixed cdm density $\Gamma$
varies by around $\pm50\%$. This is not surprising; first, most
of the microlensing is
due to objects between $10\kpc$ and $30\kpc$ from the galactic
center, so only this part of the baryonic halo is probed.
Second, the galactic halo is not well constrained by
rotation-curve data, so the ``phase space'' of reasonable models
is large \cite{ft}.

Clearly, microlensing can only provide limited
information about the local cdm density.  For example, the microlensing rate
can be as large as its value in the best fit, baryons-only
(fiducial) model in a two-component model where the local cdm density is
$3\times 10^{-25}\gcmm3$.  Or, suppose that the microlensing
rate were determined to be half the fiducial value;
for our suite of halo models the local cdm density is
$1-5\times 10^{-25}\gcmm3$.  On the other hand, if
the observed microlensing rate is found to be small, say
10\% or less of the fiducial value, based on our set
of models one could argue that the local
cdm mass density is {\it at least} $2\times 10^{-25}\gcmm3$.

To be specific about the dependence of the microlensing
rate on the halo model, for a given cdm density
the higher rates occur in models with larger baryonic core radii and
larger values of the local rotation speed.  In models with
a truncated baryonic halo (not shown) $\Gamma$ is
insensitive to the truncation radius provided that it is
greater than about $30\kpc$; this is because most of the
microlensing is due to halo objects between
$10\kpc$ and $30\kpc$ from the galactic center.  There is a
small dependence upon the distance to the LMC which is not shown; changing
the LMC distance by $\pm10\%$ changes $\Gamma$ by about $\pm5\%$.
We should emphasize that our two-component models
are very simple; one could easily imagine
more complicated models, e.g., where the halos are not spherically
symmetric.  This increases further the range of plausible
microlensing rate for a given cdm density \cite{roman}.

Finally, a fine point, in computing the microlensing rate, we have
followed Ref.~\cite{kg} in assuming that the distribution
of halo velocities is Maxwellian, which leads to
the factor of $v_H$ in $\omega_0$, cf. Eq. (\ref{eq:micro}).  This is only
strictly true for a galaxy model consisting solely of an
untruncated, zero-core radius halo.  In that regard, our models
(like most) are not self-consistent.  To explore the sensitivity of our
results to this inconsistency, we replaced $v_H$ by $\sqrt{3/2}v_{\rm circ}
(20 \kpc)$; that is, we used the circular velocity at the radius
where most of the microlensing occurs to characterize the
halo velocity dispersion.  This {\it increased} the sensitivity
of the microlensing rate to halo model, though very slightly (by
a few per cent).  Further, we also computed the optical depth
for microlensing, cf. Eq. (4) in Ref.~\cite{kg}, which does
not depend upon the distribution of halo velocities, and it exhibits
a similar variation for fixed cdm density as $\Gamma$.  Both of these facts
suggest that our estimates of the model uncertainties in the
microlensing rates---which is our main concern---have not been
affected significantly by lack of self consistency.  Absolute
rates will of course depend more strongly on consistency \cite{roman}.

\medskip
{\it Summary.}  If the density of the Universe is
significantly greater than about 10\% of critical density,
as a number of observations indicate and several arguments
suggest \cite{omega,mst1}, then there are two dark
matter problems, baryonic and
nonbaryonic.  If the nonbaryonic dark matter is cold dark matter,
which seems to be the most promising possibility, then
the halos of spiral galaxies should contain
both baryons and cdm particles.   Provided the formation
of the halo involved only gravity, the local baryonic fraction of the
halo should be small, less than about 10\%.
If baryons in the halo underwent some dissipation, so that
the baryonic halo contracted relative to the cdm halo, the
local baryonic fraction is increased by the contraction factor and could
be significantly higher.  In the extreme, if the baryonic
portion of the galaxy formed first through nongravitational
forces and a cdm halo is accreted thereafter, the mass of cdm in the halo is
comparable to that in baryons and the
local cdm density is still $10^{-25}\gcmm3$ or so.

Because of uncertainties inherent in modeling
our halo and the fact that microlensing only probes
the part of the halo between $10\kpc$ and $30\kpc$ from
the center of the galaxy, it is difficult at present to learn much about
the cdm content of own halo from microlensing.
For example, if the microlensing rate were found to
be equal to that expected in the best fit, baryons-only halo model,
the local cdm density could be as large as $3\times 10^{-25}\gcmm3$
when the uncertainties in halo models are taken into account.
On the other hand, if the microlensing rate were found to be
small, say 20\% or less of the baryons-only model, based
on our models one could argue that this is evidence for
a local cold dark matter density
larger than about $2\times 10^{-25}\gcmm3$.  While the MACHO
and EROS collaborations have yet to discuss the microlensing
rate that can be inferred from their candidate events
\cite{macho}, a naive analysis
of their data suggests that the rate could be as low as
20\% of the baryons only model \cite{mst2}

\bigskip\bigskip

We thank Josh Frieman, Kim Griest, Roman Scoccimarro and
Albert Stebbins for useful conversations and comments.
This work was supported in part by the Department of Energy
(at Chicago and Fermilab) and by the NASA through grant
NAGW-2381 (at Fermilab).

\vfill\eject
\section{Figure Captions}
\bigskip
\noindent{\bf Figure 1:}  The microlensing rate $\Gamma$ for our suite
of reasonable, two-component halo models as a function of the local cdm
density, normalized to a best fit, one-component (baryons)
halo model with core radius of $5\kpc$.  The heavy curves
show the extreme range of the models; from top to bottom the
lines correspond to models with $a_B=a_{\rm cdm}=6.5, 5, 3\kpc$.

\medskip
\noindent{\bf Figure 2:}  A sample of galactic rotation curves
for our models.  The heavy curves correspond to
extreme models (top: $a_{\rm cdm}=2\kpc$, $a_B=8\kpc$, and
$\rho_{\rm local,cdm}=0.8\times 10^{-25}\gcmm3$; bottom:
$a_{\rm cdm}=a_B=2\kpc$).  The two middle curves correspond
to $a_{\rm cdm}=a_B=6.5\kpc$ (upper), $5\kpc$ (lower).

\end{document}